\documentclass[twocolumn]{revtex4}
\usepackage{graphicx}
\usepackage{amstext}
\usepackage{amsxtra}

\begin{document}


\title{Anomalous ring-down effects and breakdown of the decay rate concept in optical cavities with negative group delay}



\author{T. Laupr\^etre$^1$}
\author{S. Schwartz$^{2}$}
\author{R. Ghosh$^3$}
\author{I. Carusotto$^4$}
\author{F. Goldfarb$^1$}
\author{F. Bretenaker$^1$}
\email{Fabien.Bretenaker@lac.u-psud.fr}
\affiliation{$^1$Laboratoire Aim\'e Cotton, CNRS-Universit\'e Paris Sud 11, 91405 Orsay Cedex, France}
\affiliation{$^2$Thales Research and Technology, Campus Polytechnique, 91767 Palaiseau Cedex, France}
\affiliation{$^3$School of Physical Sciences, Jawaharlal Nehru University, New Delhi 110067, India}
\affiliation{$^4$INO-CNR BEC Center and Dipartimento di Fisica, Universit\`a di Trento, 38123 Povo (Trento), Italy}

\date{\today}
\begin{abstract}
Propagation of light pulses through negative group velocity media is known to give rise to a number of paradoxical situations that seem to violate causality. The solution of these paradoxes has triggered the investigation of a number of interesting and unexpected features of light propagation. Here we report a combined theoretical and experimental study of the ring-down oscillations in optical cavities filled with a medium with such a strongly negative frequency dispersion to give a negative round-trip group delay time. We theoretically anticipate that causality imposes the existence of additional resonance peaks in the cavity transmission, resulting in a non-exponential decay of the cavity field and in a breakdown of the cavity decay rate concept. Our predictions are validated by simulations and by an experiment using a room-temperature gas of metastable helium atoms in the detuned electromagnetically induced transparency regime as the cavity medium.
\end{abstract}


\maketitle

\section{Introduction}
Since the early works of Sommerfeld \cite{Sommerfeld} and Brillouin \cite{Brillouin1,Brillouin2} on light propagation through resonant dielectric medium, slow and fast light have been the subject of considerable research efforts. It is now well established that the group velocity of light can change dramatically in a dispersive medium: slow, fast, or even negative group velocity light can be observed. Moreover, such effects can, under some conditions, occur without any pulse distortion \cite{Garrett}. This has led to much controversy about Einstein's causality and the propagation of a signal in such situations, which has been solved by considering the information as carried by non-analyticity points
\cite{Garrison,Kuzmich,Stenner1,Stenner2,Milonni}.

The control of group velocities of light pulses is an active subject of research as slow light schemes have been proposed to enhance nonlinear interactions for applications in quantum information processing \cite{Harris,Lukin,Lvovsky}. In recent years, the use of electromagnetically induced transparency (EIT) in high-finesse cavities has given promising results for coherent control of light and nonlinear optics at low light levels \cite{Wu, Mucke, Albert}. The question of the lifetime of the field in cavities filled with a dispersive medium has consequences also on potential applications such as the increase of the sensitivity of gyroscopes using fast light \cite{Shahriar,Salit,Smith}. In this context, we have recently confirmed experimentally that in the case of a slow-light medium inserted inside an optical cavity, the field lifetime is governed by the group velocity \cite{Laupretre}. We investigate here some paradoxes arising from the consideration of a negative group velocity medium inserted inside a cavity, and we show that in such a case one is forced to take into account the additional resonance peaks of the cavity imposed by causality, making the concept of one single cavity decay rate not relevant anymore \cite{Haroche}.

\begin{figure*}[t]
     \includegraphics[width=1.0\linewidth,angle=0]{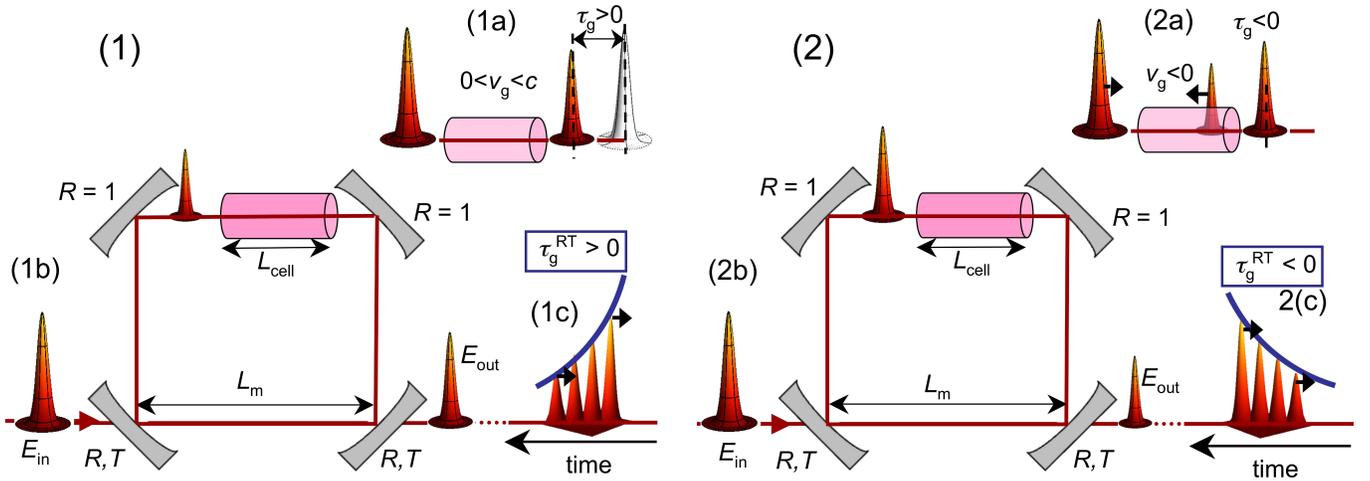}
\caption{{\bf Propagation of a light pulse} through (1a) a slow-light and (2a) negative-light medium (the white pulse in (1a) would correspond to propagation in vacuum), and (1b),(2b) a cavity containing such a medium. Series of pulses exiting (1c) the slow-light cavity when it is excited with an incident pulse, and (2c) the cavity in the case where the intracavity medium has a strong negative dispersion leading to a negative round-trip group delay $\tau_{\mathrm{g}}^{\mathrm{RT}}<0$. Notice the non-causal behavior of the exit pulses in (2c).} \label{Fig1}
\end{figure*}

Let us start by illustrating the paradox on the simplest case of a short pulse of light propagating through a dispersive
medium of refractive index $n(\omega)$ and length $L_{\mathrm{cell}}$. If the dispersion $\mathrm{d}n/\mathrm{d}\omega$ is positive (see Fig. \ref{Fig1}), the group velocity $v_{\mathrm{g}}$ is positive and the pulse experiences a
positive group delay $\tau_{\mathrm{g}}=L_{\mathrm{cell}}/v_{\mathrm{g}}$ during its propagation through the medium (see Fig.\ \ref{Fig1}(1a)). On the contrary, when the dispersion is negative enough, $v_{\mathrm{g}}$ can become negative, leading to the appearance of a negative group delay $\tau_{\mathrm{g}}$ through the medium. This can lead to the
kind of situation sketched in Fig.\ \ref{Fig1}(2a), in which the peak of the outgoing pulse leaves the medium before the incident pulse enters it while another wave packet appears at the back interface and propagates in the backward direction inside the medium \cite{Bolda}. In these paradoxical situations, causality has been shown to be ensured by the propagation of non-analyticity points at the speed $c$ of light in vacuum \cite{Stenner1,Stenner2, Milonni}.

Let us now introduce such a dispersive medium inside a resonant cavity. In the case of positive dispersion of Fig.\ \ref{Fig1}(1b), i.e., slow light, we have recently shown \cite{Laupretre} that the lifetime of the field in the cavity is given as expected by $\tau_{\mathrm{cav}}=\tau_{\mathrm{g}}^{\mathrm{RT}}/\Pi$, where $\Pi$ stands for the fractional loss per cavity round trip, and $\tau_{\mathrm{g}}^{\mathrm{RT}}=\tau_{\mathrm{g}}+L_{\mathrm{vac}}/c$ is the group delay for one round trip inside the cavity with $L_{\mathrm{vac}}$ the length of the empty part of the cavity. In this case, the reduced decay rate for the intracavity intensity can be explained in terms of a simple picture of a pulse propagating at the group velocity inside the cavity and decaying at each round trip because of losses (see the decaying pulses of Fig.\ \ref{Fig1}(1c)).

This picture does no longer hold in the case where the intracavity dispersion is negative and is strong enough not only to make the group delay across the cell $\tau_{\mathrm{g}}$ negative, but also the cavity group round-trip time $\tau_{\mathrm{g}}^{\mathrm{RT}}$ smaller than zero (Fig.\ \ref{Fig1}(2b)). In this case, if we follow the same picture as in the case of positive dispersion, the pulse that has made one round trip inside the cavity must exit the cavity before the initial pulse, and is even preceded by the pulse that has undergone two round trips inside the cavity, etc. This should lead to an increase of the intensity with time, as shown in Fig.\ \ref{Fig1}(2c), which is of course absurd. In the present article, we consider a slightly different configuration where a CW laser beam is incident on the cavity and suddenly switched off: in spite of the different time sequence, paradoxical situations appear in this case as well.

\section{Results}
\subsection{General theory}
We consider a cavity like the one in Fig.\ \ref{Fig1}. The input and output mirrors are identical, with intensity reflection and transmission coefficients given by $R$ and $T$, respectively. The two other mirrors are perfectly reflecting. We call $L_{\mathrm{m}}$ the length between the input and output mirrors. $\omega_l$ is the frequency of the laser and $\omega_p$ the considered resonant frequency of the cavity. For a generic incident excitation $E_{\mathrm{in}}(t)$, the field at the output of the cavity can be evaluated using simple linear response theory. We first consider the case of a cavity with a positive round-trip group delay $\tau_{\mathrm{g}}^{\mathrm{RT}}$. We suppose that the input laser field is monochromatic and is abruptly turned off at $t=0$. The positive-frequency part of the output field reads
\begin{equation}
\textstyle{E_{\mathrm{out}}^{(+)}(t)=\int_{-\infty}^{t}\mathrm{d}t'E_{\mathrm{in}}^{(+)}(t')\;R(t-t')}\, ,\label{eq01}
\end{equation}
where $E_{\mathrm{in}}^{(+)}(t)$ is the positive-frequency part of the input field, and $R(\tau)$ is the response function of the cavity which is zero for $\tau<0$. We can stress the fact that $R$ is causal by writing it as $R(t)=S(t)H(t)$, where $H$ is the Heavyside step function, and $S$ is the Fourier transform of the cavity transmission $\widetilde{S}(\omega)$ for a monochromatic incident field. Then Eq.\ (\ref{eq01}) simply reads
\begin{equation}
\textstyle{E_{\mathrm{out}}^{(+)}(t)=[E_{\mathrm{in}}^{(+)}\ast (SH)](t)}\, .\label{eq02}
\end{equation}
The cavity transmission for a monochromatic field of angular frequency $\omega$
is then given by
\begin{equation}
\widetilde{S}(\omega)=\frac{T
\exp{\left[\mathrm{i}\frac{\omega}{c}L_{\mathrm{m}}\right]}}{1-R
\exp{\left[\mathrm{i}\frac{\omega}{c}\left(L_{\mathrm{vac}}
+n(\omega)L_{\mathrm{cell}}\right)\right]}}\;.\label{eq03}
\end{equation}
If we suppose that $\omega$ is close to a resonance frequency
$\omega_p$ of the cavity, for which
$\exp{[\mathrm{i}\frac{\omega_p}{c}(L_{\mathrm{vac}}+n(\omega_p)L_{\mathrm{cell}})]}=1$,
then, at first order in $(\omega-\omega_p)/\omega_p$, Eq.\
(\ref{eq03}) becomes
\begin{equation}
\widetilde{S}(\omega)=\frac{T \exp{\left[\mathrm{i}\frac{\omega}{c}L_{\mathrm{m}}\right]}}
{1-R-\mathrm{i}R(\omega-\omega_p)\tau_{\mathrm{g}}^{\mathrm{RT}}}\;,\label{eq04}
\end{equation}
leading to:
\begin{equation}
\widetilde{S}(\omega)=\left(\frac{T}{R
\tau_{\mathrm{g}}^{\mathrm{RT}}}\right)\frac{\exp{\left[\mathrm{i}\frac{\omega}{c}L_{\mathrm{m}}\right]}}{\frac
{\gamma_{\mathrm{cav}}}{2}-\mathrm{i}(\omega-\omega_p)}\;,\label{eq05}
\end{equation}
where the cavity decay rate is given by
\begin{equation}
\gamma_{\mathrm{cav}}=2(1-R)/R\,\tau_{\mathrm{g}}^{\mathrm{RT}}\simeq \Pi/\tau_{\mathrm{g}}^{\mathrm{RT}}=1/\tau_{\mathrm{cav}}\, .\label{eq06}
\end{equation}
We have assumed that $1-R\ll 1$. In order to predict what a measurement of the field lifetime should give, we consider
the response of this cavity to a laser field at frequency $\omega_l$ which is turned off at $t=0$:
\begin{equation}
E_{\mathrm{in}}^{(+)}(t)=E_0 [1-H(t)]\mathrm{e}^{-\mathrm{i}\omega_l t}\;.\label{eq07}
\end{equation}
Eqs.\ (\ref{eq02}), (\ref{eq05}) and (\ref{eq07}) then lead to the output fields 
\begin{eqnarray}
 E_{\mathrm{out}}^{(+)}(t)&=&\frac{S_0 E_0\mathrm{e}^{-\mathrm{i}\omega_l(t-\frac{L_{\mathrm{m}}}{c})}}{\frac{\gamma_{\mathrm{cav}}}{2}-\mathrm{i}(\omega_l -\omega_p)}\;,
 \ \mbox{if}\; t\leq\frac{L_{\mathrm{m}}}{c}\, ,\label{eq08}\\
E_{\mathrm{out}}^{(+)}(t)&=&\frac{S_0
E_0\mathrm{e}^{-\mathrm{i}\omega_p(t-\frac{L_{\mathrm{m}}}{c})}}{\frac{\gamma_{\mathrm{cav}}}{2}-\mathrm{i}(\omega_l
-\omega_p)}\mathrm{e}^{-\frac{\gamma_{\mathrm{cav}}}{2}(t-\frac{L_{\mathrm{m}}}{c})}\;,
\nonumber\\ &&\mbox{if}\; t\geq\frac{L_{\mathrm{m}}}{c}\,
,\label{eq09}
\end{eqnarray}
with $S_0=T/R\tau_{\mathrm{g}}^{\mathrm{RT}}$, which is the standard solution for a decaying cavity, in agreement with the observations of Ref.\ \cite{Laupretre}. On the contrary, in the case of a negative light cavity for which $\tau_{\mathrm{g}}^{\mathrm{RT}}<0$,
we obtain
\begin{eqnarray}
 E_{\mathrm{out}}^{(+)}(t)&=&\frac{S_0 E_0(\mathrm{e}^{-\mathrm{i}\omega_l(t-\frac{L_{\mathrm{m}}}{c})}-\mathrm{e}^{ -(\mathrm{i}\omega_p+\frac{\gamma_{\mathrm{cav}}}{2})
 (t-\frac{L_{\mathrm{m}}}{c})})}{\frac{\gamma_{\mathrm{cav}}}{2}-\mathrm{i}(\omega_l-\omega_p)}\;,\nonumber\\ &&\mbox{if}\; t\leq\frac{L_{\mathrm{m}}}{c}\, ,\label{eq10}\\
  E_{\mathrm{out}}^{(+)}(t)&=&0\;, \ \mbox{if}\; t\geq\frac{L_{\mathrm{m}}}{c}\, ,\label{eq11}
\end{eqnarray}
which, once again, clearly violates causality.

In deriving Eqs.\ (\ref{eq10}) and (\ref{eq11}), the only hypothesis that we have made is that the cavity transmission could be reduced to a single Lorentzian peak (see Eqs.\ (\ref{eq03}) and (\ref{eq04})). This hypothesis is valid as long as the spectrum of the incident field (given by Eq.\ (\ref{eq07})) is contained in a single cavity transmission peak and all frequencies experience the same group index.

\begin{figure}[t]
     \includegraphics[width=1.0\linewidth]{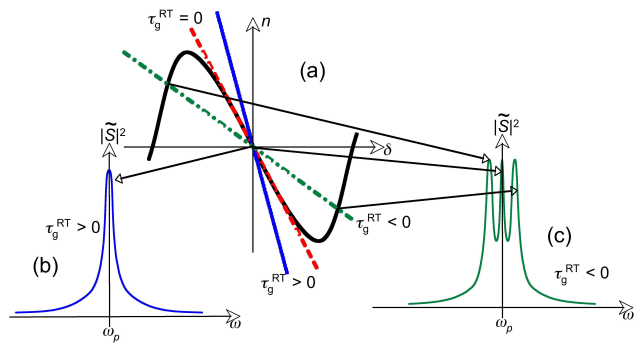}
 \caption{(a) Continuous black line: typical negative dispersion curve. The intersections of this curve with the red dashed ($\tau_{\mathrm{g}}^{\mathrm{RT}}=0$), blue dotted ($\tau_{\mathrm{g}}^{\mathrm{RT}}>0$), and green dot-dashed ($\tau_{\mathrm{g}}^{\mathrm{RT}}<0$) curves determine whether the resonance is (b) single peaked or (c) multi-peaked. 
 }\label{Fig2}
\end{figure}

In order to examine this condition, let us first consider, as an example, a typical negative dispersion curve as given by the continuous line in Fig.\ \ref{Fig2}(a). Let us suppose, without any loss of generality, that the inflection point of the dispersion curve occurs at the empty cavity resonance frequency $\omega_p$, meaning that $\frac{\omega_p}{c}
(L_{\mathrm{vac}}+n(\omega_p)L_{\mathrm{cell}})=2 p \pi$, where $p$ is an integer. Let us try to determine whether extra resonance peaks, due to negative dispersion, could occur in the vicinity of the peak at $\omega_p$. If $\omega_p+\delta$ is the angular frequency of such an extra peak, the resonance condition reads
\begin{equation}
\frac{\omega_p+\delta}{c} (L_{\mathrm{vac}}+n(\omega_p+\delta)L_{\mathrm{cell}})=2 p \pi\,.\label{eq12}
\end{equation}
To first order in $\delta/\omega_p$, this condition is equivalent to
\begin{equation}
n(\omega_p+\delta)-n(\omega_p)=-\frac{L_{\mathrm{vac}}+n(\omega_p)L_{\mathrm{cell}}}{\omega_p L_{\mathrm{cell}}}\delta \,.\label{eq13}
\end{equation}

The left-hand side of Eq.\ (\ref{eq13}) versus $\delta$ is the continuous line in Fig.\ \ref{Fig2}(a). The right-hand side is a straight line, as shown by the dotted, dashed, and dot-dashed lines in Fig.\ \ref{Fig2}(a). One can see that the shape of the resonance, namely, the existence of no other solution than $\delta=0$, leading to a single peak as in Fig.\ \ref{Fig2}(b), or the existence of two other resonance frequencies for $\delta \neq 0$, leading to two
extra resonance peaks as in Fig.\ \ref{Fig2}(c), depends on the relative values of the slopes of the dispersion curve and the line corresponding to the right-hand side of Eq.\ (\ref{eq13}). In particular, the condition for the existence of two extra solutions reads, at first order in $\delta$:
\begin{equation}
\left. -\frac{\mathrm{d}n}{\mathrm{d}\omega}\right|_{\omega_p}>\frac{L_{\mathrm{vac}}+n(\omega_p)L_{\mathrm{cell}}}{\omega_p L_{\mathrm{cell}}}\, ,\label{eq14}
\end{equation}
which is equivalent to $\tau_{\mathrm{g}}^{\mathrm{RT}}<0$. We thus reach the following conclusion: the fact that the group delay for one round trip inside the cavity is negative leads to the existence of satellite peaks around the resonance considered. This negates the approximation used to obtain the cavity transmission (see Eq.\ (\ref{eq04})) and explains why the non-causal situation described above can actually never be reached. Figure \ref{Fig2}(c) illustrates how this condition results in the existence of two extra peaks for the cavity resonance labeled by the integer $p$. Note that in the case of slow light,  the slope of the dispersion curve in Fig.\ \ref{Fig2}(a) would be reversed, allowing only one intersection with the continuous line and thus forbidding the existence of extra resonance peaks.

\begin{figure}[h]
     \includegraphics[width=1\linewidth]{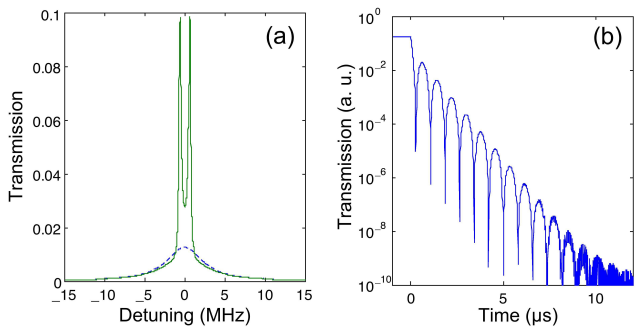}
 \caption{(a) Theoretical cavity transmission versus detuning in the presence of (continuous green line) and without (blue dashed line) an intracavity gain doublet creating negative light. (b) Corresponding decay of the intracavity intensity when the incident field is turned off at $t=0$. With the parameters used, the group delay for one round trip inside the cavity is $\tau_{\mathrm{g}}^{\mathrm{RT}}=-3.3\, \mathrm{ns}$.
 }\label{Fig3}
\end{figure}

In reaching the conclusion above, we have assumed a specific form for the negative dispersion curve, namely, the one drawn in Fig.\ \ref{Fig2}(a). We will now show that this conclusion holds, as a consequence of causality, for any arbitrary negative dispersion curve provided $\tau_{\mathrm{g}}^{\mathrm{RT}}<0$ at $\omega_p$.
For this, it is convenient to introduce the following function:
\begin{equation}
f(\delta)=L_{\mathrm{cell}} [(\omega_p+ \delta)
n(\omega_p+\delta)-\omega_p n(\omega_p)] + \delta L_{\mathrm{vac}}
\,.\nonumber
\end{equation}
It can then be shown, by a straightforward calculation, that Eq.\ (\ref{eq12}) is equivalent to $f(\delta)=0$. The resonance peak at $\omega_p$ corresponds to $f(0)=0$, while the satellite peaks would correspond to non-zero solutions of $f(\delta)=0$. The initial hypothesis $\tau_{\mathrm{g}}^{\mathrm{RT}}<0$ at $\omega_p$ is equivalent to $\partial f /\partial \delta<0$ for $\delta = 0$. Moreover, for $\delta \rightarrow +\infty$, causality imposes that $n(\omega) \rightarrow 1$ \cite{Jackson}, which results in $f \sim (L_\mathrm{vac} + L_\mathrm{cell})\delta$. To summarize, we have
$f(\delta)=0$ and $\partial f/\partial \delta <0 $ for $\delta=0$. So $f(\delta)$ has to go to negative values for small (strictly positive) values of $\delta$. Also, since $f(\delta) \rightarrow + \infty$ for $\delta \rightarrow +\infty$, it has to go at least once through zero, according to the intermediate value theorem. Let $\delta_1$ be the smallest value for which this happens. Then $\omega_p+\delta_1$ corresponds to an additional resonance peak of the cavity. Similarly, it is possible to prove the existence of $\delta_2 <0$ such that $f(\delta_2)=0$, which corresponds to another additional resonance peak. This leads to the conclusion that at least two satellite peaks exist around $\omega_p$ for any arbitrary dispersion curve as soon as $\tau_{\mathrm{g}}^{\mathrm{RT}}<0$ at $\omega_p$.

\subsection{Example of a gain doublet}
Let us be more specific about the situation in which a medium can exhibit a strong negative dispersion. A very popular example of negative dispersion is provided by a gain doublet \cite{Steinberg,Wang,Dogariu,Kuzmich,Stenner1,Pati}. Figure
\ref{Fig3}(a) shows the transmission $|\widetilde{S}(\omega)|^2$ of the cavity versus detuning in that case. The dashed curve corresponds to the empty cavity, which is 2.45~m long with 29\% losses per round trip. We now suppose that a gain-doublet medium is inserted inside the cavity. The two gain peaks are separated by 1.5\ MHz. We suppose that they are located symmetrically with respect to the cavity resonance. The gain maxima correspond to 28\% per round trip and the full width at half maximum of each peak is 800~kHz. In these conditions, the group delay for one round trip inside the
cavity is $\tau_{\mathrm{g}}^{\mathrm{RT}}=-3.3\, \mathrm{ns}$. The corresponding intensity transmission spectrum of the cavity is reproduced as a continuous line in Fig.\ \ref{Fig3}(a). One can clearly see the two transmission peaks corresponding to the conjugated effects of the two additional resonance peaks and of the two gain maxima. The main difference with respect to Fig.\ \ref{Fig2}(c), which was computed using only the real part of the dispersion and by artificially setting the imaginary part to zero (no gain or absorption) is that there is no central transmission
peak. This is consistent with the fact that there is no gain peak at zero detuning. One can also notice in this spectrum that the two lateral peaks are slightly shifted towards the line center with respect to the positions of the atomic resonances, which is consistent with the fact that these gain peaks are located in a positive dispersion spectral region.

We calculate (using Eqs.\ (\ref{eq02}), (\ref{eq03}) and (\ref{eq07})) the temporal evolution of the intensity
$|E_{\mathrm{out}}^{(+)}(t)|^2$ at the output of the cavity when the incident field is suddenly turned off at $t=0$. Such a decay is represented in Fig.\ \ref{Fig3}(b) on a logarithmic scale. It is clearly non-exponential. It consists of a fast decay by two orders of magnitude, followed by oscillations which correspond to beatnotes between the two peaks of the transmission spectra. It is an illustration of the general principle of Fig.\ \ref{Fig2}: any intracavity negative dispersion effect which is strong enough to make the round-trip group delay negative will cause secondary transmission peaks to emerge that will make the cavity decay non-exponential, forbidding one to define a field lifetime for this
cavity.

\begin{figure}[ht]
     \includegraphics[width=1\linewidth]{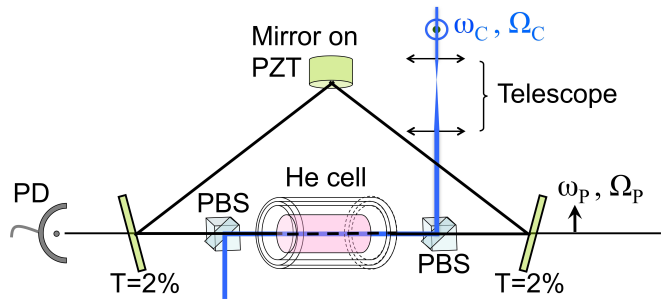}
 \caption{Experimental set-up. PBS: polarization beam splitter. PZT: piezoelectric transducer. PD: photodetector. The cell is protected from spurious magnetic fields by a mu-metal shield.} \label{Fig4}
\end{figure}

\subsection{Experiments with detuned EIT}
In order to give an experimental illustration, we use another system in which a large negative group delay can be achieved: detuned EIT in a hot vapor of metastable $^4$He atoms \cite{Goldfarb}. We use a 6-cm long cell filled with 1 Torr of helium at room temperature. Some of these atoms are excited to the $^3$S$_1$ metastable state using an RF discharge at 27\,MHz. Metastable helium is well known for exhibiting a pure three-level $\Lambda$ system when excited at the 1.083\,$\mu$m transition between the $2^3$S$_1$ and $2^3$P$_1$ energy levels using circularly polarized light. Light at 1.083\,$\mu$m is provided by a single-frequency diode laser. The frequencies and Rabi frequencies of the coupling and probe beams used in our experiment are driven by two AOMs. A telescope expands the coupling beam diameter up to 0.5~cm, which is larger than the probe beam diameter. The cell is inserted inside a 2.4-m long triangular ring cavity made of two plane mirrors with 2\% transmission and a high reflectivity concave mirror with a 5-m radius of curvature. The cavity is resonant only for the probe field as two polarization beam-splitters drive the coupling beam inside and outside the cavity \cite{Laupretre} (see Fig.\ \ref{Fig4}).

\begin{figure}[ht]
     \includegraphics[width=1\linewidth]{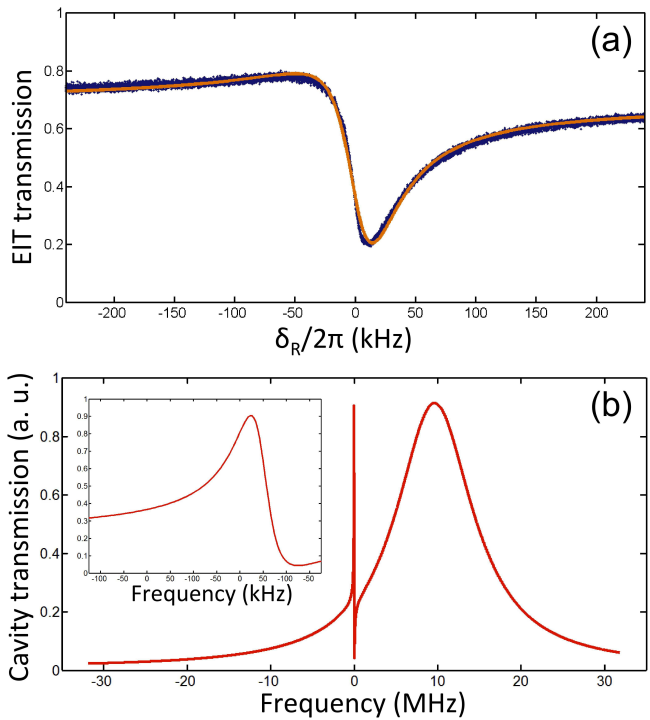}
 \caption{(a) Experimental (thick blue:) transmission of the cell without optical cavity versus Raman detuning $\delta_R$ with an optical detuning $\Delta_c$ around 1.4~GHz and fit (thin orange) using the theoretical expressions of Ref. \cite{Goldfarb}. (b) Calculated cavity transmission profile as a function of the probe frequency. The frequency reference is taken at $\delta_R=$~0. Inset: zoom around the narrow resonance.} \label{CavityEIT}
\end{figure}

Let us first consider the transmission of our degenerate three-level $\Lambda$ system taken outside the cavity. A large one-photon detuning transforms the transparency peak typical of EIT into an asymmetric absorption peak \cite{Mikhailov}. Consequently, in the vicinity of the transmission minimum, the system exhibits strong negative dispersion, which leads to a negative group delay. Indeed, if we introduce such an optical detuning $\Delta_c$ between the coupling field and the maximum of the Doppler profile of the transition, the evolution of the cell transmission versus Raman detuning $\delta_R$ between the two fields exhibits the asymmetric Fano-like profile shown in Fig.\ \ref{CavityEIT}(a), which is fitted using the expressions of the susceptibility derived in Ref. \cite{Goldfarb} for a three-level system in detuned EIT. With a coupling power of 5~mW equivalent to a coupling Rabi frequency around 10~MHz, and an optical detuning $\Delta_c$ around 1.4~GHz, we measure a negative group delay $\tau_{\mathrm{g}}\approx -4\, \mu\mathrm{s}$ around the absorption maximum ($\delta_R=5\ \mathrm{kHz}$) for our 6-cm long cell when the cavity is not present. The measurement is made by modulating the probe signal amplitude at a 1~kHz frequency, so that the spectrum of the field is fully contained in the absorption dip of a few kHz width.

\begin{figure}[t]
     \includegraphics[width=1\linewidth]{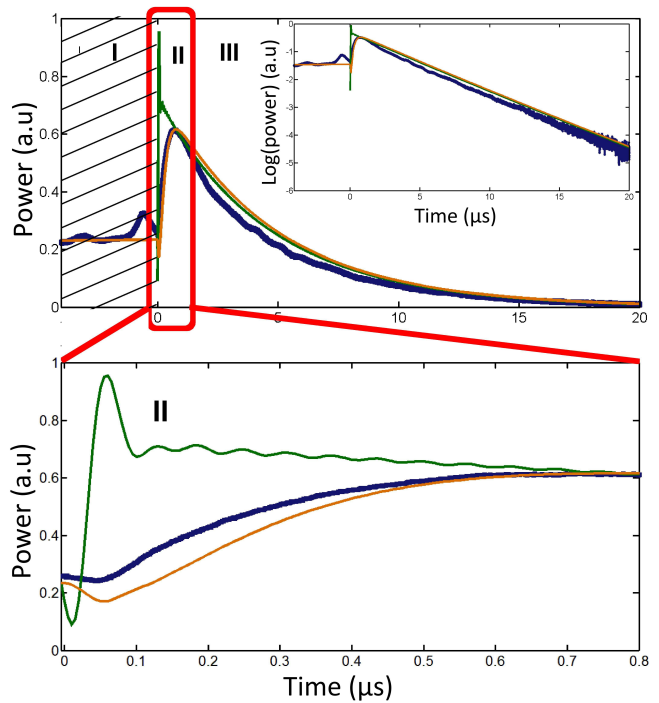}
 \caption{Thick blue: Experimental decay of the intracavity intensity when the incident probe field is turned off at $t=0$. Thin green: Corresponding theoretical cavity decay. Orange dashed: Corresponding theoretical cavity decay when the 200 ns response time of the AOM is taken into account in the model. Inset: Same plot on a logarithmic scale. Below: Zoom on the first part of the evolution of the experimental decay of the intracavity intensity when the incident probe field is turned off at $t=0$.} \label{ResultsEIT}
\end{figure}

Once this cell is inserted inside the cavity, we apply the same coupling field to the atoms. The probe field, which is then slightly detuned from the coupling field ($\delta_R=5\ \mathrm{kHz}$) in order to take advantage of the large negative group delay of -4~$\mu$s, is incident on the cavity input mirror. We slowly scan the length of the cavity using a piezoelectric actuator that carries one of the mirrors. When the cavity is at resonance with the probe, i.e., when the cavity transmission reaches a given threshold, we abruptly turn off the probe field using an AOM. We then observe the evolution of the intensity at the output of the cavity.

The thick blue curve in Fig.\ \ref{ResultsEIT} is the experimentally recorded evolution of the intensity at the output of the cavity when we abruptly turn off the probe field. In this figure, region labeled I shows the signal before switching off the cavity input field. The fluctuations are due to cavity length instabilities. The interesting part of the signal lies on region labeled II, that shows that the intensity starts increasing, on a time scale shorter than 1~$\mu$s, before decreasing. This evolution is clearly non-exponential, showing once more that a negative cavity round-trip group delay leads to a non-exponential decay of the intracavity field. Notice here that the cavity length being equal to 2.4~m, the cavity round-trip group delay $\tau_{\mathrm{g}}^{\mathrm{RT}}$ is also of the order of $ -4\, \mu\mathrm{s}$.

These experimental results are consistent with the theoretical calculations (based on Eqs.\ (\ref{eq02}), (\ref{eq03}) and (\ref{eq07})) using the expression for inhomogeneously broadened detuned EIT as in Ref.\ \cite{Goldfarb}. The values of the parameters have been obtained by fitting the experimental transmission curve as we can see on Fig.\ \ref{ResultsEIT}. With a coupling Rabi frequency of 11~MHz, an optical detuning $\Delta_c=1.3\ \mathrm{GHz}$ and a Raman coherence decay rate of $12\,\mathrm{kHz}$, we obtain the thin green curve of Fig.\ \ref{ResultsEIT}, which is similar to the experimental result (thick blue): when the incident intensity is turned off, the intensity at the output of the cavity starts increasing before decreasing. This typical non-exponential decay constitutes one more illustration of the behavior described in Fig.\ \ref{Fig2}.

The small difference between experimental and theoretical results can be explained by four immediate reasons: i) The photodiode used to detect the signal has a time response of the order of 200~ns, which does not allow us to detect the fast variations of the signal; ii) the time taken to turn off the probe field is not zero but driven by the fall time of the acousto-optic modulator (AOM), which is smaller than the photodiode response time; iii) the extra losses of the polarization optics have been set to 1~\% in the theoretical model and can be slightly different in the experiment; iv) before the probe signal is shut off the signal is a bit noisy due to the fluctuations of the laser frequency and of the cavity length, but this does not play any role in the further evolution of the cavity decay as soon as the incident signal is cut off. In order to check the influence of the fall time of the AOM, we introduce its response in our model by replacing the Heavyside function for the incident field by an exponential decay with a 200~ns decay time. This leads to the orange dashed curve of Fig.\ \ref{ResultsEIT}, which is in very good agreement with our experimental results.

Because the temporal response we obtain is directly linked to the spectral response of the cavity, it is possible to qualitatively understand the origin of the different time scales involved. Figure \ref{CavityEIT}(b) reproduces the calculated cavity transmission spectrum, taking into account the negative group velocity medium of our experiment. Of course, narrow spectral features will drive slow temporal behaviors, while on the contrary broad spectral features will be related to fast temporal behaviors. In the particular experimental situation we present in this paper, the spectral response of the cavity exhibits a single narrow resonance. So we can expect, as observed experimentally and confirmed by theoretical results, an exponential decay driven by a large time constant at the end of the temporal behavior. This can be seen in the region labeled III and the inset of Fig.\ \ref{ResultsEIT}. The inset of Fig.\ \ref{CavityEIT}(b) shows a zoom on this narrow resonance. Its width is approximately equal to 40~kHz, which corresponds to the 4~$\mu$s time constant of the exponential decay shown in region III of Fig.\ \ref{ResultsEIT}. In the experiment, we fill the cavity with a light frequency for which the transmission is very weak ($\delta_R=5\ \mathrm{kHz}$). By turning off the incident laser very fast, we excite some frequencies for which the cavity is much more resonant. So when the cavity empties, it starts by reaching a light intensity level larger than the one previously reached in the steady-state regime during its excitation by a monochromatic CW beam. We can then expect an initial fast increase of the signal as shown in the zoomed part of Fig.\ \ref{ResultsEIT}. Moreover, the small oscillations with a time period of a few tens of ns that can be seen in the theoretical curve correspond to the beatnote between the narrow resonance and the wide resonance of the cavity transmission profile, which are separated by roughly 9~MHz. In the experimental profile, the oscillations are smoothed out because of the finite response time of the acousto-optic switch, as reproduced by the orange dashed line in Fig.\ \ref{ResultsEIT}.

\section{Discussion}
In conclusion, we have theoretically demonstrated the fact that it is impossible to obtain a negative group delay for one round trip inside a resonant cavity while keeping an exponential decay of the intracavity intensity. We have shown that this result is a consequence of causality in negative group-delay cavities. We have illustrated this both numerically and experimentally, by using negative velocity light induced by a gain doublet and detuned EIT in a metastable vapor, respectively. Our result is consistent with the fact that slow light is usually associated with a transparency peak, which reduces the bandwidth to be considered. On the contrary, fast and negative group velocity light appears in the case of an absorption peak, leading to the possibility of many frequencies playing a role, and thus the cavity decay rate in such a cavity can no longer be simply defined. This should have interesting consequences on the spontaneous emission rate of atoms placed in such a cavity \cite{Bradshaw}, with application to the spontaneous emission noise of lasers based on such negative light cavities.

\begin{acknowledgments}
The authors acknowledge partial support from the Agence Nationale de
la Recherche, the Triangle de la Physique, and the Indo-French Center
for the Promotion of Advanced Research (IFCPAR/CEFIPRA). IC
acknowledges financial support from ERC through the QGBE grant.
\end{acknowledgments}

\end{document}